# IS BOHR'S CHALLENGE STILL RELEVANT?

**Leonardo Chiatti**
AUSL VT Medical Physics Laboratory
Via Enrico Fermi 15, 01100 Viterbo (Italy)
fisica1.san@asl.vt.it

**Summary**

We argue that not all the theoretical content of the Bohr model has been captured by the "definitive" quantum formalism currently in use. In particular, the notion of "quantum leap" seems to refer to non-dynamic features, closely related to non-locality, which have not yet been formalized in a satisfactory way.

**Introduction**

The Bohr-Rutherford planetary model [1] is still the general public's favourite image of the mystery of the atom because of its simplicity and "visualizability". For students of physics and chemistry it represents a sort of inevitable rite of passage on the path towards orbitals and quantum mechanics. As an educational tool, it allows the soft introduction of the quantum of action $h$, by its appearance in a series of constraints on the otherwise classical motion of electrons represented as classical material points.
Even in a scientific perspective, the model has partly re-emerged in the context of semi-classical approaches to the quantization of atomic and molecular structures [2,3]. In particular, the references [4,5,6] show that by removing the condition of the impenetrability of matter (this assumption seems reasonable with reference to the time when the model was developed, but is now known to be less significant on a microphysical scale) originally introduced by Bohr and Sommerfeld, a different counting of states is obtained, in better agreement with that offered by quantum mechanics.
On the centenary of its formulation (1913) it is appropriate to consider whether, beyond its educational role, the Bohr model is still relevant to current research on the foundations of quantum mechanics. This short paper argues that it is. This personal centennial celebration does not propose any new concept, but nevertheless covers a range of issues that are generally overlooked in the historical and educational debate on the Bohr model.

**Context**

Firstly, it is worth mentioning that the Bohr model was not the first atomic model to include the quantum of action $h$[1]. After Planck's seminal work, Johannes Stark was probably the first physicist to understand the link between this new constant and the micro-world [7,8,9], playing an important role in the dissemination of this concept among German physicists in the first decade of the 20th century[2].
In 1910, Arthur Erich Haas presented his quantum model of the hydrogen atom [11,12,13],

---

[1] In this paper we do not discuss J.W. Nicholson's important contribution.
[2] Although, oddly enough, he did not include this concept in the atomic model [10] he proposed.

probably as a result of these suggestions. In contemporary terms, it could be said that Haas derived a semi-classical quantization of the ground state of this atom, in the context of Thomson's plum-cake model. This approach provided the correct expression for the radius of the atom in terms of the charge and mass of the electron and $h$ (currently known as "the Bohr radius"). However, it is not clear whether Haas considered $h$ as a new fundamental constant. His choice of Thomson's model as the theoretical framework was largely justified by its classical stability (unlike Rutherford's planetary model). The sole purpose of quantization was here to constrain the radius.

It is a well known fact that Bohr chose Rutherford's model as his framework; indeed, his assiduous presence in Rutherford's laboratory allowed him to acquire first-hand results of well known experiments that led to the rejection of Thomson's model. Further, Bohr was firmly convinced of the fundamental nature of the quantum of action [1].

The existence, fundamentality and irreducibility of this quantum make the analysis of physical phenomena over time, with a level of detail equivalent to variations of action significantly smaller than $h$ impossible. Thus it is reasonable to expect processes that cannot be causally analyzed on an atomic scale. Bohr was aware of the need for a new mechanical approach (based on the finiteness of $h$) expressing this limitation, and of the purely provisional nature of a representation of the atom based on electronic orbits and other classical elements.

**Quantum leaps**

Putting a part their different theoretical framework, which is irrelevant here, Bohr's model contributed three important innovative elements compared to Haas. It leads:

1) to the quantization of the motion of material bodies (quantization of orbits)

2) to the field quantization (the application of Planck's law to emitted/absorbed radiation: the concept of *photon*)

3) to the existence of discontinuous leaps from one stationary state to another (quantum leaps)

Haas' proposal led uniquely to result 1), limited exclusively to the ground state, and did not address transitions – points 2) and 3). Bohr's model therefore allowed predictions about the position of spectral lines that were impossible with Haas' approach.

The subsequent developments are discussed in textbooks on the history of physics [14,15]. The research inspired by Bohr's model paved the way for the construction of a complete system of formally self-consistent quantum physics. With regards to point 1), this led to quantum mechanics (QM) and the first quantization formalism, whereas point 2) led to quantum field theory (QFT) and the second quantization formalism. The history books, as well as the accounts of the legacy of the Bohr model which is currently considered entirely absorbed in QM and QFT formalisms, normally stop here.

However, point 3) also exists. In experiments involving micro-objects, a "quantum leap" is the event which prepares the initial quantum state, or detects the final quantum state. In other words, quantum leaps are somehow connected to the projection of the temporal evolution of the initial state on the final state: the infamous "collapse" of the wave function. This collapse is controversial as QM formalism does not specify when and how this event occurs. This lack of formal description, according to a minimalist interpretation such as the Copenhagen interpretation, leads to well-known paradoxes such as Wigner's friend, Schrödinger's cat, and so forth.

Therefore it can safely be asserted that point 3) has not yet been properly developed in the context of an appropriate self-consistent formalism. The challenge launched by Bohr is still relevant to this day.

**Beyond time**

Given the above, it is reasonable to ask why Bohr decided to frame QM in a meta-theoretical structure – the well-known "Copenhagen interpretation", designed to suppress a priori any question about the effective location and structure of quantum leaps. Indeed, for a long time, these issues were well beyond any permissible limits.

A possible answer is given by examining the mechanics of "leaps" in Bohr's original model. One problem arises immediately: given that transitions are only permitted between allowed levels, how does an electron know whether there is a free level to leap onto and thus take flight? Once the electron has taken a leap, how does it manage to end its leap exactly on the required level without trial and error[3]?

It should be emphasised that even current quantum formalism fails to provide a comprehensive answer to this problem. In this formalism, the electronic orbital is a superposition of the initial and final orbitals, with time-dependent coefficients obtained by solving the time-dependent Schrödinger equation. Although the probability of the electron being located in the final state increases with time, the electron is always – at any given moment – either on the initial or the final orbital. Indeed, a measure projecting the electronic state onto these orbitals will always yield one of these two outcomes.

Although measuring the electron's position repeatedly on a *ensemble* of identical preparations shows the gradual evolution of the probability distribution of its location to that corresponding to the final orbital, the transition of single atom occurs at a definite point in time, characterised by the emission/absorption of a photon.

This implies the sudden transition (in the case of a single atom) of an entire extended orbital and raises issue of non-locality. However, this non-locality is confined to the atom and the moment of the leap and is not related to the connection between different events. Consequently, it is not the same non-locality as that of entangled states, for example. Nevertheless the existence of this "hidden" non-locality is the modern version of the ancient objection to the Bohr model.

This problem can be avoided by assuming that the quantum leap is not a dynamic process. To clarify, suppose there is an a-spatial and a-temporal physical reality, simply referred to as *background*. A quantum leap can then be modelled as a double dissolution process in this *background* of the physical state of a micro-object, for example an atomic state, followed by the emergence of a new physical state. Since the *background* is a-temporal, this dual process has no duration and from an observer's (*foreground*) perspective, the quantum leap is instantaneous. The two physical states connected by the leap are not causally related in the sense of temporal dynamics represented by differential equations, but rather in the sense of an eternal and universal algebra of states. Thus, the problem of a non-local causal evolution (even though in the "hidden" and innocuous sense described above) is bypassed.

The introduction of a synchronous correlation between *background* and *foreground* restricted to quantum leaps is arguably a totally unnecessary dialectic game, since the non-locality of quantum leaps is "internal" and therefore quite innocuous. However, the approach adopted for atomic leaps can also be applied to systems of entangled particles undergoing to measurements by distant observers; non-locality becomes, in this case, evident. It enables this type of situation to be comfortably addressed, allowing the source of non-locality to be identified in the *background*.

---

[3] Rutherford's objection to Bohr, referred for example in [16].

## Theories of nothing

Thus, quantum leap structure theory requires the definition of a theoretical framework for the physical observables emerging from the *background*, and their re-absorption into it. Such a description must be non-dynamic and therefore based on algebra and logic rather than differential equations.

It should be noted that this approach must include spatial and temporal position as observables; thus, contrary to popular belief, the emergence of the spatial-temporal order should be defined on an atomic/particle scale, and not necessarily on the Planck scale. A second important observation is that this type of approach should constrain possible physical states (and interactions) starting from non-dynamic general conditions and should therefore be "archetypal", in the philosophical sense. These archetypal conditions should define, for example, the spectrum of elementary particles and their interactions. Basic interactions are in fact expressed as quantum leaps in the *foreground*, and particle states are specific connections between quantum leaps.

Paradoxically, the outcome is a "theory of nothing" rather than a "theory of everything", because the inevitable starting point would be an adequate definition of the *background*; and from a *foreground* perspective, *background* is pure nothing.

There are several historical examples of branches of theories that tended towards this type of approach (but remained incomplete) including, among others: Von Weizsäcker's [17] qubit approach; Bohm and Hiley's [18,19] holoalgebra (and the holomovement); Finkelstein's space-time code [20, 21, 22, 23]; Rowlands' [24] Universal Rewrite System; Pierre Noyes' [25, 26] bit string physics; and Stuckey and Silberstein's [27] "block world" QFT. In a sense, even Chew and Capra's [28, 29] latest version of bootstrap belongs to this group, although this approach focuses on archetypal constraints for matrix S rather than on quantum leaps. Even Schroer's [30] "algebraic" QFT has many points of similarity with these research programmes, although remaining quite distinct.

## Conclusions

The Bohr intransigent promotion and defence of the "Copenhagen interpretation" probably resulted from the need to avoid a premature contact of physics community with the a-temporal aspect, without adequate theoretical tools to assimilate it in the context of physical theory.

Without these tools, we would be dealing with a purely qualitative reference to a vague concept with "mystical" objective connotations, and this could prevent a newly introduced theory from being accepted by the scientific community. Conceptual problems were averted by relegating them to the background, and the focus was on applications with productive outcomes: the entire micro-world had to be discovered.

Small groups of researchers finally managed to bring unresolved issues to the fore [31,32] only after the thrill of exploring new territories had passed. This led to some important discoveries, including the non-locality of quantum theory.

In this centenary of the Bohr model, which introduced the concept of "quantum leap" explicitly for the first time, we may ask whether the time has come for a more in-depth investigation of this mysterious process.